\documentclass[12pt]{iopart}
\usepackage[top=2.5cm,bottom=2.5cm,left=2.5cm,right=2.5cm]{geometry}
\usepackage{graphicx}
\usepackage{bm}
\usepackage{epsfig}
\usepackage{import}

\expandafter\let\csname equation*\endcsname\relax
\expandafter\let\csname endequation*\endcsname\relax
\usepackage{amsfonts} 
\usepackage{amsmath}
\usepackage{iopams}
\usepackage{amssymb}
\usepackage{mathtools}
\usepackage{cancel} 
\usepackage{bm}
\usepackage[usenames, dvipsnames]{color} 
\usepackage{caption}
\usepackage{dcolumn}
\usepackage{epic} 
\usepackage{epsfig}
\usepackage{feynmp}
\usepackage{graphicx}
\usepackage{grffile}
\usepackage[breaklinks,colorlinks = true,linkcolor = red,urlcolor=blue,citecolor=red]{hyperref}

\usepackage{mathrsfs}
\usepackage{soul}
\usepackage{wrapfig}
\usepackage{xy} 
\usepackage{xcolor}
\usepackage{amsmath,amssymb,amsfonts,amsthm}
\usepackage[ascii]{inputenc}
\usepackage{makecell}
\allowdisplaybreaks

\hypersetup{
    colorlinks=true,       
    linkcolor=black,          
    citecolor=blue,        
    filecolor=magenta,      
    urlcolor=cyan           
}

\newcommand{\be}{\begin{equation}}
	\newcommand{\ee}{\end{equation}}
\newcommand{\ba}{\begin{aligned}}
	\newcommand{\ea}{\end{aligned}}
\newcommand{\bea}{\begin{eqnarray}}
	\newcommand{\eea}{\end{eqnarray}}	
	
\newcommand{\beal}{\begin{align}}
\newcommand{\eal}{\end{align}}

\usepackage{fancyhdr}
\begin{document}

\title{Relating absorbing and hard wall boundary conditions for a one-dimensional run-and-tumble particle}

\author{Mathis Gu\'eneau}
\address{Sorbonne Universit\'e, Laboratoire de Physique Th\'eorique et Hautes Energies, CNRS UMR 7589, 4 Place Jussieu, 75252 Paris Cedex 05, France}
\ead{mathis.gueneau@protonmail.com}
\author{L\'eo Touzo}
\address{Laboratoire de Physique de l'Ecole Normale Sup\'erieure, CNRS, ENS and PSL Universit\'e, Sorbonne Universit\'e, Universit\'e Paris Cit\'e,
24 rue Lhomond, 75005 Paris, France}
\ead{leo.touzo@phys.ens.fr}



\begin{abstract}

The connection between absorbing boundary conditions and hard walls is well established in the mathematical literature for a variety of stochastic models, including for instance the Brownian motion. In this paper we explore this duality for a different type of process which is of particular interest in physics and biology, namely the run-tumble-particle, a toy model of active particle. For a one-dimensional run-and-tumble particle subjected to an arbitrary external force, we provide a duality relation between the exit probability, i.e. the probability that the particle exits an interval from a given boundary before a certain time $t$, and the cumulative distribution of its position in the presence of hard walls at the same time $t$. We show this relation for a run-and-tumble particle in the stationary state by explicitly computing both quantities. At finite time, we provide a derivation using the Fokker-Planck equation. All the results are confirmed by numerical simulations.
\end{abstract}

\maketitle
\tableofcontents
\hypersetup{linkcolor=red}

\section{Introduction}
Recently, there has been a surge of interest in first passage properties of active particles. Inspired from biological systems such as bacteria, active particles exhibit a persistent motion. Unlike their passive counterparts, such as Brownian motion, they can convert energy from their environment into motion, and are thus inherently out-of-equilibrium \cite{activeintro1,activeintro2,activeintro3,activeintro4,activeintro5}. Recently, active particles have been the focus of extensive research, leading to the development of various models like the active Ornstein-Uhlenbeck particle (AOUP) \cite{Wijland21, AOUP}, the active Brownian particle (ABP) \cite{activeintro4,ABM}, or the run-and-tumble particle (RTP) \cite{ Tailleur_RTP, Cates2012} (also called persistent random walk \cite{Kac1974, Ors1990, PRWWeiss}). This last process is inspired from the motion of \textit{Escherichia coli} \cite{Berg2004} and consists in a series of straight runs over exponentially distributed times separated by instantaneous tumbles, during which the particle takes a random orientation (recent experiments also suggest other distributions for the run times, such as a power law \cite{natureruntime} in which case these RTPs are similar to L\'evy walks \cite{levywalks}).
When confined within a specific geometry, or inside an external potential, due to this persistence, active particles tend to accumulate at the boundaries of the accessible domain \cite{Lee2013, Yang2014, Uspal2015, Duzgun2018, AngelaniHardWalls, hardWallsJoanny, hardWallsCaprini}. Specifically, the distribution of the position of a one-dimensional RTP inside a confining potential has been observed to display a non-Boltzman steady state \cite{DKM19, Sevilla,Velocity_RTP}, transitioning from a passive-like regime to an active-like regime depending on the persistence time \cite{DKM19}.

Characterizing the first passage properties \cite{redner, Metzler_book, EVS1} of active particles is of significant importance. For instance, understanding how small organisms navigate toward targets like food, or how sperm seek an egg cell presents a substantial challenge \cite{benichou1,benichou2, Targetsearch}. Extracting these properties for active particles proves immensely difficult, primarily due to the persistence of their motion stemming from the colored nature of their stochastic noise -- being correlated in time, unlike white noise \cite{colorednoise}. In the case of a free RTP in 1d, the survival probability and mean first time (MFPT) have been computed in various settings \cite{Velocity_RTP,MalakarRTP,SurvivalRTPDriftDeBruyne, Singh2020,Singh2022,MFPT1DABP}. Recently, the MFPT of a one-dimensional RTP in confining potentials has been explicitly calculated, revealing that the MFPT can be minimised with respect to the tumbling rate $\gamma$ \cite{MFPT_1D_RTP}. Other studies focus on a free RTP in confined domains with various boundary conditions, in one dimension \cite{Angelani1,BressloffStickyBoundaries,RTPpartiallyAbsorbingTarget} or higher \cite{TVB12, RBV16,RTPsurvivalMori}. A related quantity is the exit probability (or splitting/hitting probability \cite{MalakarRTP,handbookSM,hittingProbaAnomalous}). Imagine two absorbing walls located at $x=a$, and $x=b$ (where $b>a$), and suppose that the RTP initiates its motion at some position $x$ in $[a,b]$. What is the probability that the particle is first absorbed at the right wall $x=b$ ? The finite time counterpart of this quantity (i.e. the probability of reaching $x=b$ before time $t$) is key as it contains all first passage properties of the system. Indeed, if one knows the probability that the particle exits at $x=a$
and the probability that it exits at $x=b$ at any time, one can deduce the survival probability at all times. 

\begin{figure}[t]
\centering
    \begin{minipage}[c]{1\linewidth}
        \centering
        \includegraphics[width=1.\linewidth]{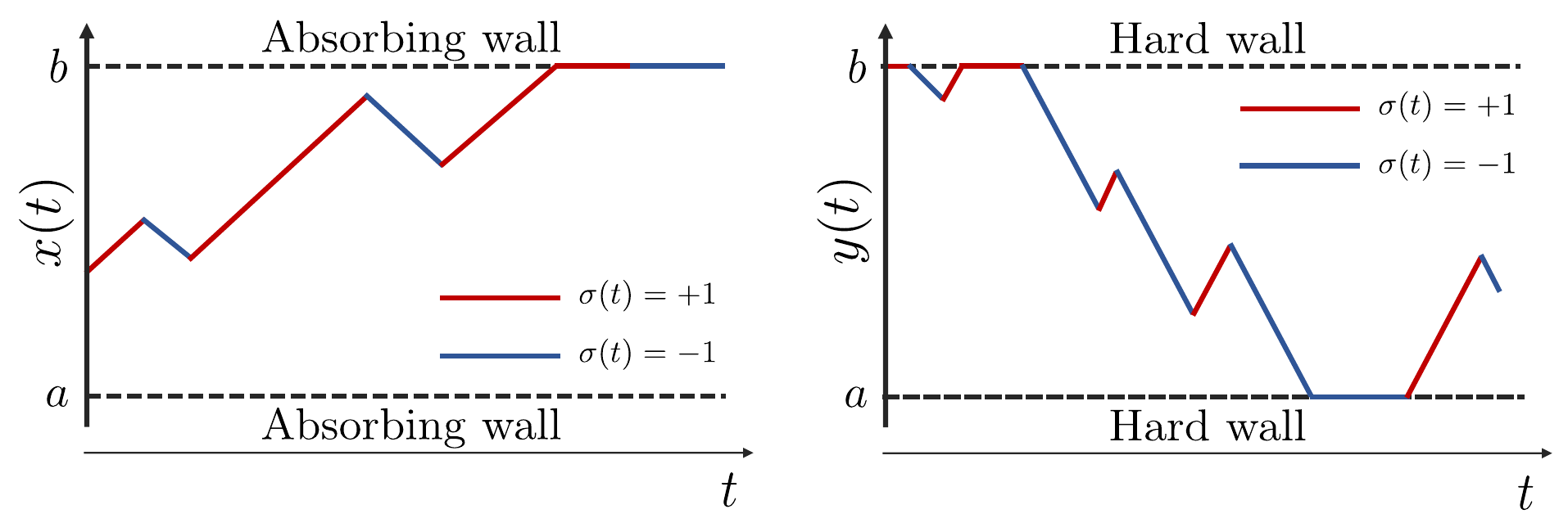}
    \end{minipage}
    \caption{We consider an RTP that evolves through the equation of motion $\dot{x}(t)=F(x) + v_0\, \sigma(t)$, where $\sigma(t)$ is a telegraphic noise switching from values $+1$ to $-1$ at exponentially distributed times. On the left panel we show a schematic trajectory of an RTP with absorbing walls located at $a$ and $b$. When the RTP reaches one of these walls, it will stay there forever no matter its state $\sigma=\pm 1$. On the right panel, we show a trajectory of another RTP $y(t)$, with hard walls (i.e. if the particle reaches one of the walls and tries to cross it, it stays in the same place).}
\label{figureRTPwalls}
\end{figure}

In this paper we consider a run-and-tumble particle subjected to an arbitrary external potential, with two types of boundary conditions, represented in Figure \ref{figureRTPwalls}: absorbing walls and hard walls. When a particle reaches an \textit{absorbing wall} it remains stuck to it indefinitely. Conversely, a {\it hard wall} can be considered as an infinite potential step, i.e. when a particle encounters it, it remains there until its total velocity changes sign. This is sometimes called a ``reflecting'' boundary condition, but we prefer the terminology ``hard wall'' in the context of active particles, to make it clear that the particle does not instantly tumble when it reaches the wall and can remain stuck for a certain time do to the persistence in its motion.
The aim of this paper is to highlight the surprising connection which exists between these two types of boundary conditions, in the context of active particles. More precisely, the exit probability of a run-and-tumble particle can be related to its cumulative distribution in the presence of hard walls. This kind of relation has been shown for various types of models in the mathematical literature, where it is known as Siegmund duality \cite{Levy,Lindley,Siegmund}. Two stochastic processes, $x(t)$ and $y(t)$, are said to be Siegmund duals if they satisfy the condition at any time $t$,
\begin{equation} \label{Siegmund_general}
    \mathbb{P}(x(t) \geq y | x(0)=x) = \mathbb{P}(y(t) \leq x | y(0)=y)\, .
\end{equation}
Siegmund proved the existence of a Siegmund dual for any stochastically monotone Markov process, i.e. when $\mathbb{P}(x(t) \geq y|x(0) = x)$ is a non-decreasing function of $x$ \cite{Siegmund}. However, the explicit construction of the dual is not trivial in general. 

In this paper we will be focusing on the particular case where $y=b$. In that case, the left-hand side of \eqref{Siegmund_general} coincides with the exit probability (since when $x(t)$ reaches $b$ it stays there at any later time). Consider for instance a Brownian motion $x(t)$ inside a potential $V(x)$ with absorbing boundary conditions at $a$ and $b$. After an infinite time, the probability that $x(t)$ exits the interval through $x=b$ is given by $\int_a^x dz\, e^{\frac{V(z)}{D}} / \int_a^b dz\, e^{\frac{V(z)}{ D}}$ (where $D$ is a diffusion constant). This probability is also the cumulative distribution with hard walls at $a$ and $b$, in the stationary state, if the potential is $-V(x)$ (as noticed for instance in \cite{hittingProbaAnomalous}). Therefore, the dual process, $y(t)$, has identical dynamics to $x(t)$ but with hard walls at $a$ and $b$ and reversed potential $-V(x)$. One can show that this duality also holds at finite time, as given by \eqref{Siegmund_general}.

Active particle models, while non-Markovian when considering only the particle's position, become Markovian when considering both position and driving velocity. In fact, the existence of a dual was also shown for processes driven by a stationary process, in both discrete and continuous time \cite{AsmussenDiscrete,SigmanContinuous}. The applications considered were however mostly oriented towards mathematical finance (the exit probability is called ``ruin probability'' in this context).

Siegmund duality is a special case of Markov duality, which has been applied in various fields such as queuing theory, finance, population genetics, and interacting particle systems (see e.g. \cite{JansenDualityReview} for a general review). However, this duality relation might not be as widely recognized within the physics community, even though some similar relations have been observed \cite{hittingProbaAnomalous, Comtet2011, Comtet2020, ThibautDual}. This paper explores this duality for the run-and-tumble particle, a celebrated model of active particles.
\bigskip

\bigskip

We first derive in Sec.~\ref{exitprobaRTPsubsection} the exit probability at infinite time for an RTP with an arbitrary external force. This computation was performed before in \cite{MalakarRTP} but only for a free RTP (although with an additional Brownian noise). In Sec.~\ref{cumulativeRTPstationarySection}, we then calculate the cumulative distribution of the position of an RTP when the walls are hard walls (drawing inspiration from \cite{AngelaniHardWalls,DKM19}). We clarify the connection that exists between the two quantities in Sec.~\ref{SectionRTPDUAL}. This relation is however not restricted to the stationary state, and in Sec.~\ref{sec:finite_time} we show that it holds at any time, given the right initial conditions. The case where the force has turning points (i.e. points where the external force is stronger than the velocity of the RTP, making some regions of space inaccessible to the particle) is slightly more delicate and we treat it in a separate section, Sec.~\ref{generalforceRTP}. We conclude in Sec.~\ref{discussion} with a discussion of the implications of this duality.

Although we focus here on the case of RTPs, this duality relation is actually much more general. Indeed, one can show that it holds not only for other well-known models of active particles such as AOUPs and ABPs, but also for other families of models such as diffusing diffusivity \cite{DiffDiffChubynsky, DiffDiffFPTSposini} or stochastic resetting \cite{resettingPRL, resettingReview}, and even some instances of discrete and continuous time random walks (see \cite{LongPaper}).

\section{Exit probability of a run-and-tumble particle}\label{exitprobaRTPsubsection}

We consider an RTP subjected to a force $F(x)$, that may derive from a potential $V(x)$ through the relation $F(x) = -V'(x)$. The dynamics is as follows,
\begin{equation}
    \dot x(t)=F(x) + v_0 \, \sigma(t)\, ,
\label{langeRTP}
\end{equation}
where $v_0$ is the inner speed of the particle. The telegraphic noise $\sigma(t)$ switches value at rate $\gamma$ through 
\begin{eqnarray}
\sigma(t+dt) = \begin{cases}
\sigma(t) &\text{, with probability } (1 - \gamma \, dt)\\
-\sigma(t) &\text{, with probability } \gamma\, dt 
\end{cases}\, .
\label{telegraphicdynamics}
\end{eqnarray}
The transitions between these values are referred to as \textit{tumbles}. The time duration $\tau$ between two consecutive tumbles follows an exponential distribution with a probability density function given by $p(\tau) = \gamma\, e^{-\gamma \tau}$. When the RTP is in the positive state $\sigma = +1$, we describe the particle as being in a ``positive" ($+$)  state, whereas when it is in the negative state $\sigma = -1$, we refer to it as being in a ``negative" ($-$) state. Its initial position is $x(0)=x\in [a,b]$ and two absorbing walls are located at $x=a$ and $x=b$ (see Figure \ref{figureRTPwalls}). In this section, we will assume that $|F(x)| < v_0$ on the interval $[a,b]$ such that the whole interval is accessible to the particle no matter its starting position. When the force has turning points $F(x) = \pm v_0$, the behaviour of the particle is more subtle. We comment on such cases in Section \ref{generalforceRTP}. 

We consider the probability for the RTP to exit at wall $b$, before or at time $t$, with a positive (resp. negative) initial velocity
\begin{equation}
    E_b(x,\pm,t) = \mathbb{P}\left(x(t)=b\, |\, x(0)=x, \sigma(0)=\pm 1 \right)\, .
\end{equation}
Recall that when the particle reaches $b$, it stays there forever -- see Fig. \ref{figureRTPwalls}. In this section we will be focusing on the infinite time limit
\begin{equation}
    E_b(x,\pm) = \lim_{t \to +\infty} E_b(x,\pm,t) \;,
\end{equation}
i.e. we want to know the probability that after an infinite time, the particle has exited at $b$ and not at $a$. If we assume $\sigma(0) = +1$ or $\sigma(0)=-1$ with probability $1/2$, the exit probability of an RTP regardless of the initial speed is
\begin{equation}
    E_b(x) = \frac{1}{2}\left(E_b(x,+)+E_b(x,-)\right)\,.
\end{equation} 
We can write a pair of coupled first-order differential equations for $E_b(x,\pm)$. Let us evolve the particle during the time interval $[0,dt]$ and average over the possible trajectories. Suppose the RTP starts its motion at $x$, then for the positive state of the RTP, the only possible events are that the particle switches sign, with probability $\gamma\, dt$, and moves from $x$ to $x + \left[F(x)-v_0\right]\, dt$, or that it moves over a distance $\left[F(x)+v_0\right] \, dt$ while staying in the positive state, with probability $1-\gamma dt$. This translates to
\begin{equation}
    E_b(x,+,t+dt)=(1-\gamma dt)\, E_b(x+\left[F(x)+v_0\right]\, dt,+, t) + \gamma dt E_b(x+\left[F(x)-v_0\right]\, dt,-, t)\, .
\end{equation}
After Taylor-expanding at order $dt$ (and repeating the operation for $E_b(x,-,t)$), we obtain
\begin{eqnarray}\label{timeBFPplus}
    \partial_t E_b(x,+,t)=\left[F(x)+v_0\right]\partial_x E_b(x,+,t) +\gamma E_b(x,-,t)- \gamma E_b(x,+,t)\, , \\
    \partial_t E_b(x,-,t)=\left[F(x)-v_0\right]\partial_x E_b(x,-,t) +\gamma E_b(x,+,t)- \gamma E_b(x,-,t)\, , \label{timeBFPminus}
\end{eqnarray}
In the stationary state, this becomes
\begin{eqnarray}
\left[F(x)+v_0\right] \partial_x E_b(x,+) + \gamma\, E_b(x,-) -\gamma\,  E_b(x,+) &=& 0\, , \label{RTP2gamma1}\\
\left[F(x)-v_0\right] \partial_x E_b(x,-) + \gamma\, E_b(x,+) - \gamma\, E_b(x,-) &=& 0 \label{RTP2gamma2}\, .
\end{eqnarray}
To solve the coupled equations (\ref{RTP2gamma1}) and (\ref{RTP2gamma2}), one needs two boundary conditions. When an RTP starts in the negative state at position $a$, it is absorbed at the wall $a$ and thus never reaches $b$ leading to $E_b(a,-)=0$. On the other hand, if an RTP starts in the positive state at wall $b$ it will directly exit the interval such that $E_b(b,+) = 1$.

Writing equations (\ref{RTP2gamma1})-(\ref{RTP2gamma2}) in terms of $E_b(x)=\frac{1}{2}\, (E_b(x,+)+E_b(x,-))$ and $e_b(x)=\frac{1}{2}\, (E_b(x,+)-E_b(x,-))$ we get
\begin{eqnarray}
&&F(x)\,  \partial_x E_b(x) + v_0\,  \partial_x e_b(x) = 0\, ,\label{sumdifeq1} \\
&&F(x)\,  \partial_x e_b(x) + v_0\,  \partial_x E_b(x)  - 2\gamma \, e_b(x) =0\, .\label{sumdifeq2}
\end{eqnarray}
Using equation (\ref{sumdifeq1}), we can replace $E_b(x)$ in equation (\ref{sumdifeq2}) to get a first order equation for $e_b(x)$, which we solve. From this we deduce $E_b(x)$, using $E_b(a,-)=0$ and $E_b(b,+) = 1$ to fix the integration constants. We obtain
\begin{eqnarray} \label{exitprobaexplicitRTP} 
    \!\!\!\!\!\!\!\!\!\!\!\!\!\!\!\!\!\!\!  &&E_b(x) = \frac{1}{Z} \left(2\gamma v_0 \int_{a}^x dz\, \frac{\exp \left[ -2\gamma \int_{a}^z du\, \frac{F(u)}{v_0^2 - F(u)^2} \right]}{v_0^2-F(z)^2} + 1 \right)\, , \\
    \!\!\!\!\!\!\!\!\!\!\!\!\!\!\!\!\!\!\!  &&Z = 2\gamma v_0 \int_{a}^b dz\, \frac{\exp \left[ -2\gamma \int_{a}^z du\,  \frac{F(u)}{v_0^2 -F(u)^2} \right]}{v_0^2-F(z)^2} + \exp \left[ -2\gamma \int_{a}^b du\, \frac{F(u)}{v_0^2 - F(u)^2} \right] + 1\, ,
\end{eqnarray}
and using that $E_b(x,\pm) = E_b(x) \pm e_b(x)$, we have
\begin{equation}
E_b(x,\pm) = \frac{1}{Z} \left( 2\gamma v_0 \int_{a}^x dz\, \frac{\exp \left[ -2\gamma \int_{a}^z du\, \frac{F(u)}{v_0^2 - F(u)^2} \right]}{v_0^2-F(z)^2} \pm \exp \left[ -2\gamma \int_{a}^x du \frac{F(u)}{v_0^2 - F(u)^2} \right] + 1 \right).
\label{exprEpm_final}
\end{equation}

\textit{Diffusive limit:}
An RTP behaves as a diffusive particle when taking the limit $\gamma \to +\infty$, and $v_0 \to \infty$, while keeping the ratio $D_a = v_0^2/(2\gamma)$ fixed (where the subscript `a' stands for active). In this diffusive limit, using Eqs.~(\ref{exitprobaexplicitRTP}), and if one assumes that the force derives from a potential $F(x)=-V'(x)$, we recover the well known result
\begin{equation}
E_b(x) = \frac{\int_{a}^x dz \, e^{\frac{V(z)}{D_a} }}{\int_{a}^b dz \, e^{\frac{V(z)}{D_a} }}\, .
\end{equation}
\textit{Free RTP:}
When there is no force applied to the RTP, the particle is free and $F(x) = 0$. In this case, it is quite straightforward to compute the exit probabilities. From Eq. (\ref{exitprobaexplicitRTP}) and Eq.~(\ref{exprEpm_final}), one obtains
\begin{equation}
    E_b(x) = \frac{\frac{1}{2}+\frac{\gamma}{v_0}(x-a)}{1+\frac{\gamma}{v_0}(b-a)}\, , \quad E_b(x,+) = \frac{1+\frac{\gamma}{v_0}(x-a)}{1+\frac{\gamma}{v_0}(b-a)}\, ,  \quad {\rm and} \quad  E_b(x,-) = \frac{x-a}{\frac{v_0}{\gamma}+(b-a)}\, .
\end{equation}
The result is linear, as in the Brownian case where the exit probability is simply $(x-a)/(b-a)$.

{\it Constant drift :} The exit probabilities $E_b(x,\pm)$ (\ref{exprEpm_final}) can be computed exactly for a range of potentials. Let us start with the simplest case, namely a constant drift $F(x)=\alpha$ with $|\alpha|<v_0$. One finds
\begin{eqnarray}
&&E_b(x,\pm) = \frac{1}{Z} \left[ \left(1+\frac{v_0}{\alpha}\right) + \left(\pm 1 - \frac{v_0}{\alpha}\right) e^{-\frac{2\gamma \alpha}{v_0^2-\alpha^2}(x-a)} \right] \, ,\\
&&Z = 1 + \frac{v_0}{\alpha} + \left(1 - \frac{v_0}{\alpha}\right) e^{-\frac{2\gamma \alpha}{v_0^2-\alpha^2}(b-a)}\, .
\label{constantdrifteq}
\end{eqnarray}

{\it Harmonic potential :} It is also possible to obtain an explicit form for the exit probabilities of an RTP inside a harmonic potential $V(x)= \mu\, x^2 / 2$. The force is $F(x) = -\mu\, x$ and we impose $b<v_0/\mu$ and $a> -v_0/\mu$ such that $|F(x)|<v_0$ in $[a,b]$. The exit probabilities are given by (for any $x$)
\begin{equation}\label{explicithramonicEb}
\begin{split}
    \!\!\!\! E_b(x,\pm)=\frac{1}{Z}\left\{1 \pm \left(\frac{{v_0^2 - a^2 \mu^2}}{{v_0^2 - \mu^2 x^2}}\right)^{\frac{\gamma}{\mu}} + \frac{2\, \gamma}{v_0} \left(1 - \frac{a^2 \mu^2}{v_0^2}\right)^{\frac{\gamma}{\mu}} \left[x \, _2F_1\left(\frac{1}{2}, 1+\frac{\gamma}{\mu}, \frac{3}{2}, \frac{\mu^2 x^2}{v_0^2}\right) \right.\right.\\ 
    \left.\left.- a \, _2F_1\left(\frac{1}{2},1+ \frac{\gamma}{\mu}, \frac{3}{2}, \frac{a^2 \mu^2}{v_0^2}\right)\right]\right\}\, ,
\end{split}
\end{equation}  
where $_2F_1$ is the hypergeometric function.
The expression of the normalisation constant is
\begin{equation}
\begin{split}
      \!\!\!\!\!\!\!\!\!\!\!\! Z = 1 + \left(\frac{{v_0^2 - a^2 \mu^2}}{{v_0^2 - b^2 \mu^2}}\right)^{\frac{\gamma}{\mu}} +
\frac{2\, \gamma}{v_0} \left(1 - \frac{a^2 \mu^2}{v_0^2}\right)^{\frac{\gamma}{\mu}}\left[b \, _2F_1\left(\frac{1}{2}, 1 + \frac{\gamma}{\mu}, \frac{3}{2}, \frac{b^2 \mu^2}{v_0^2}\right) \right. \\
\left.- a \, _2F_1\left(\frac{1}{2}, 1+ \frac{\gamma}{\mu}, \frac{3}{2}, \frac{a^2 \mu^2}{v_0^2}\right)\right]  \, .
\end{split}
\end{equation}
One can check from Eq.~(\ref{explicithramonicEb}) that $E_b(x,-)$ vanishes linearly as $x\to a^+$, and that $E_b(b^-,-)<1$ (and similarly $E_b(a^+,+)>0$, and $1-E_b(x,+)$ vanishes linearly as $x\to b^-$).

\begin{figure}[t]
\centering
    \begin{minipage}[c]{.49\linewidth}
        \centering
        \includegraphics[width=1.\linewidth]{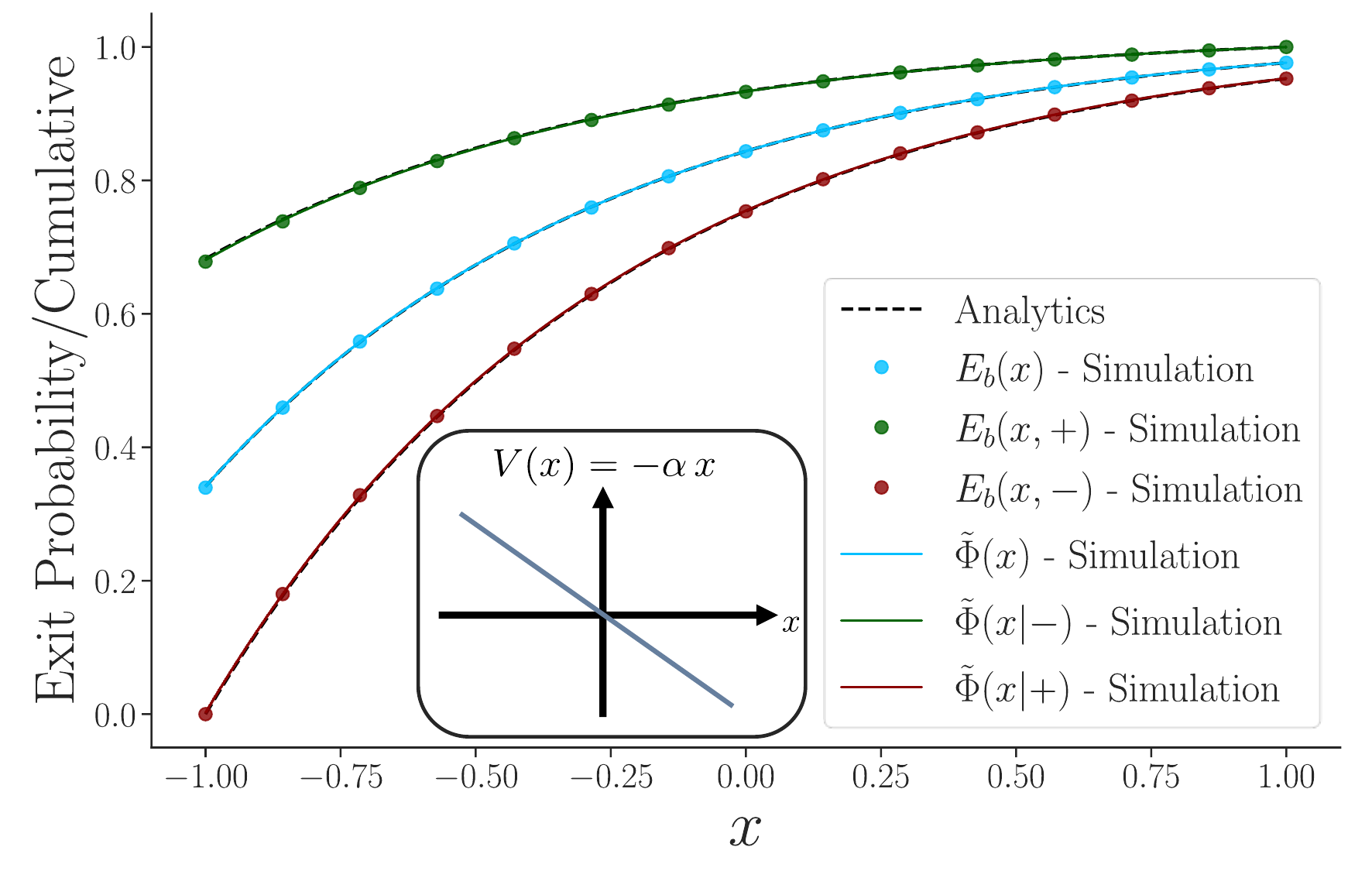}
    \end{minipage}
    \hfill%
    \begin{minipage}[c]{.49\linewidth}
        \centering
        \includegraphics[width=1.\linewidth]{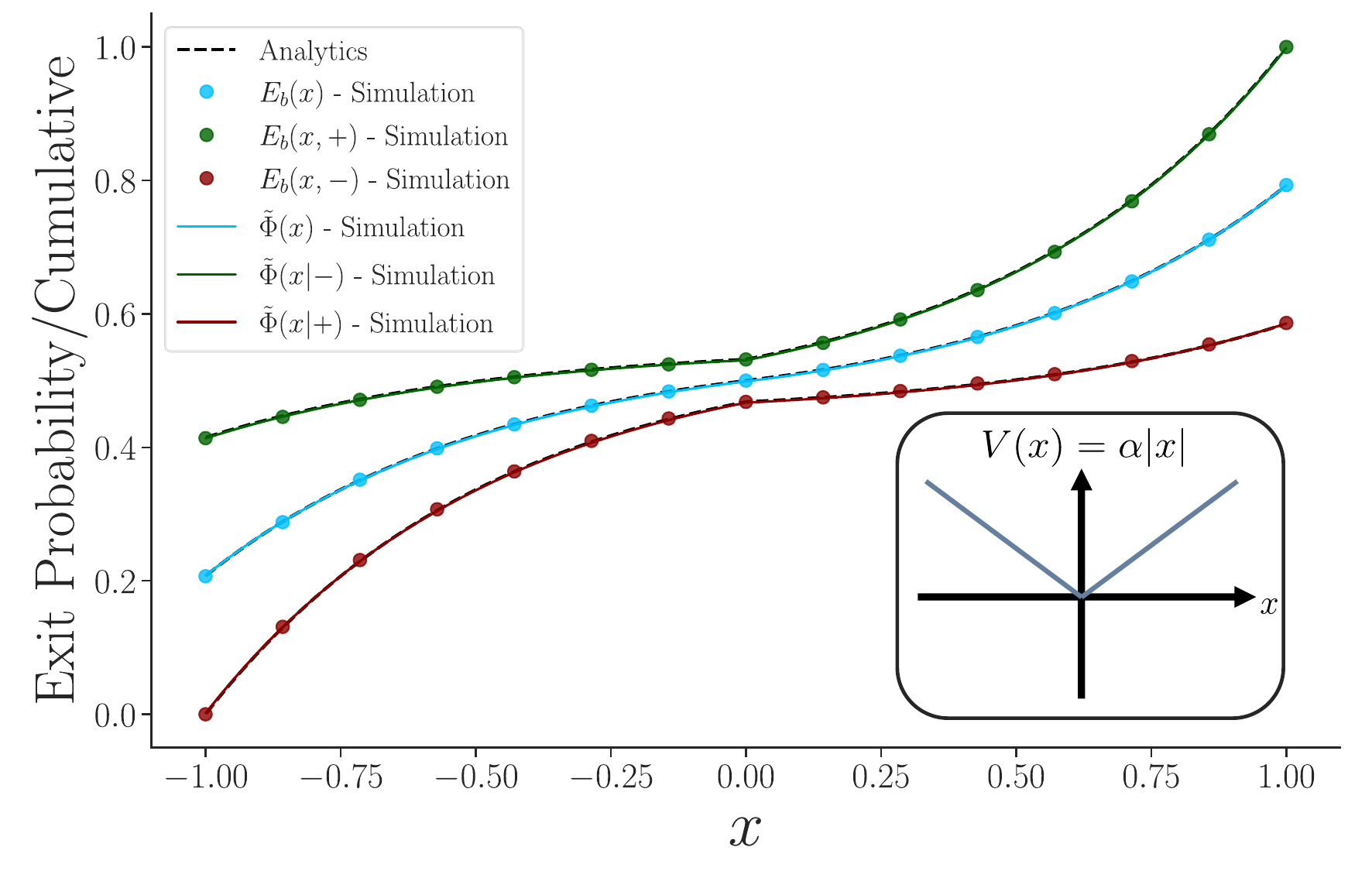}
    \end{minipage}
    \hfill%
    \begin{minipage}[c]{.49\linewidth}
        \centering
        \includegraphics[width=1.\linewidth]{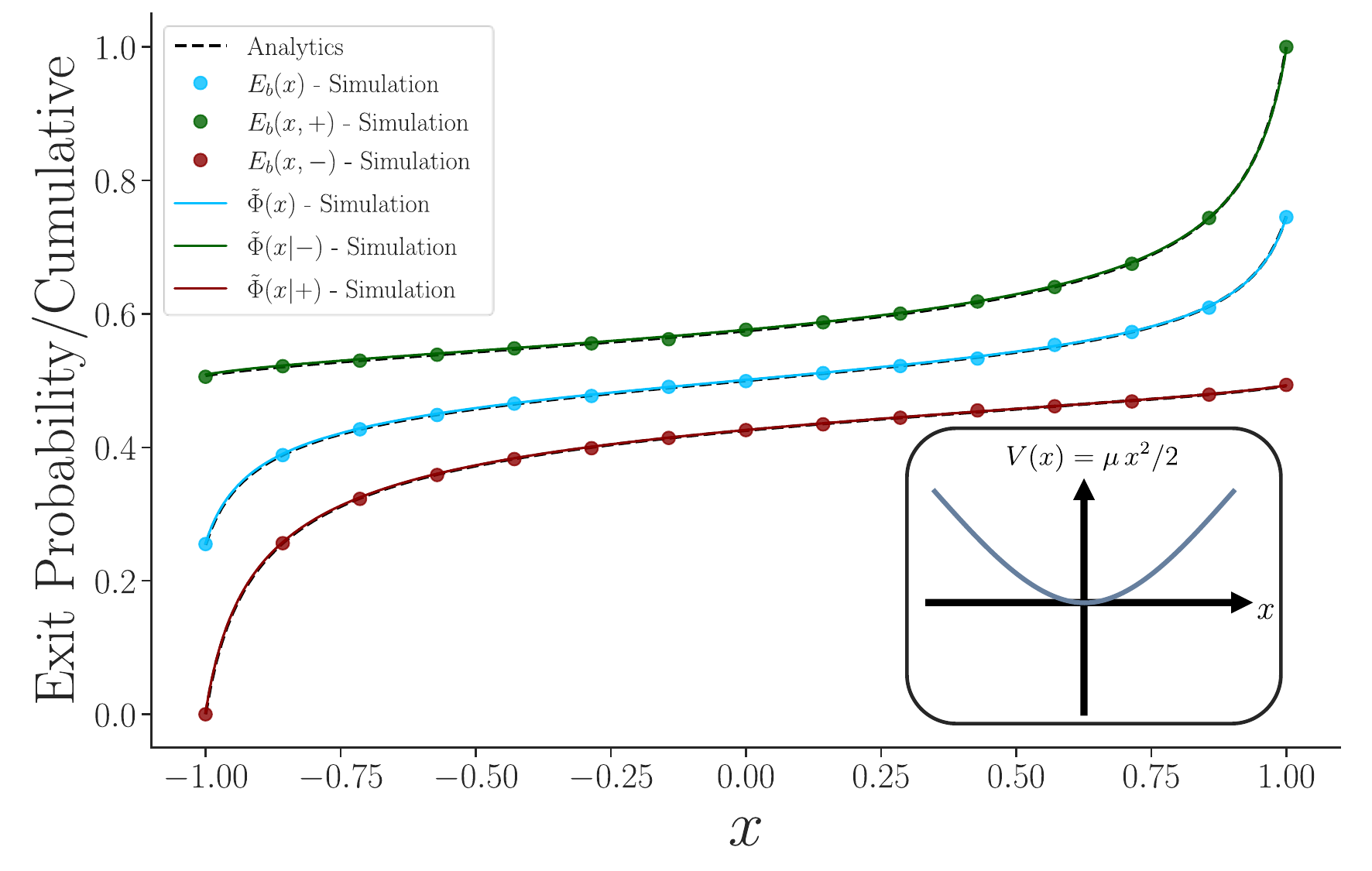}
    \end{minipage}
    \hfill%
    \begin{minipage}[c]{.49\linewidth}
        \centering
        \includegraphics[width=1.\linewidth]{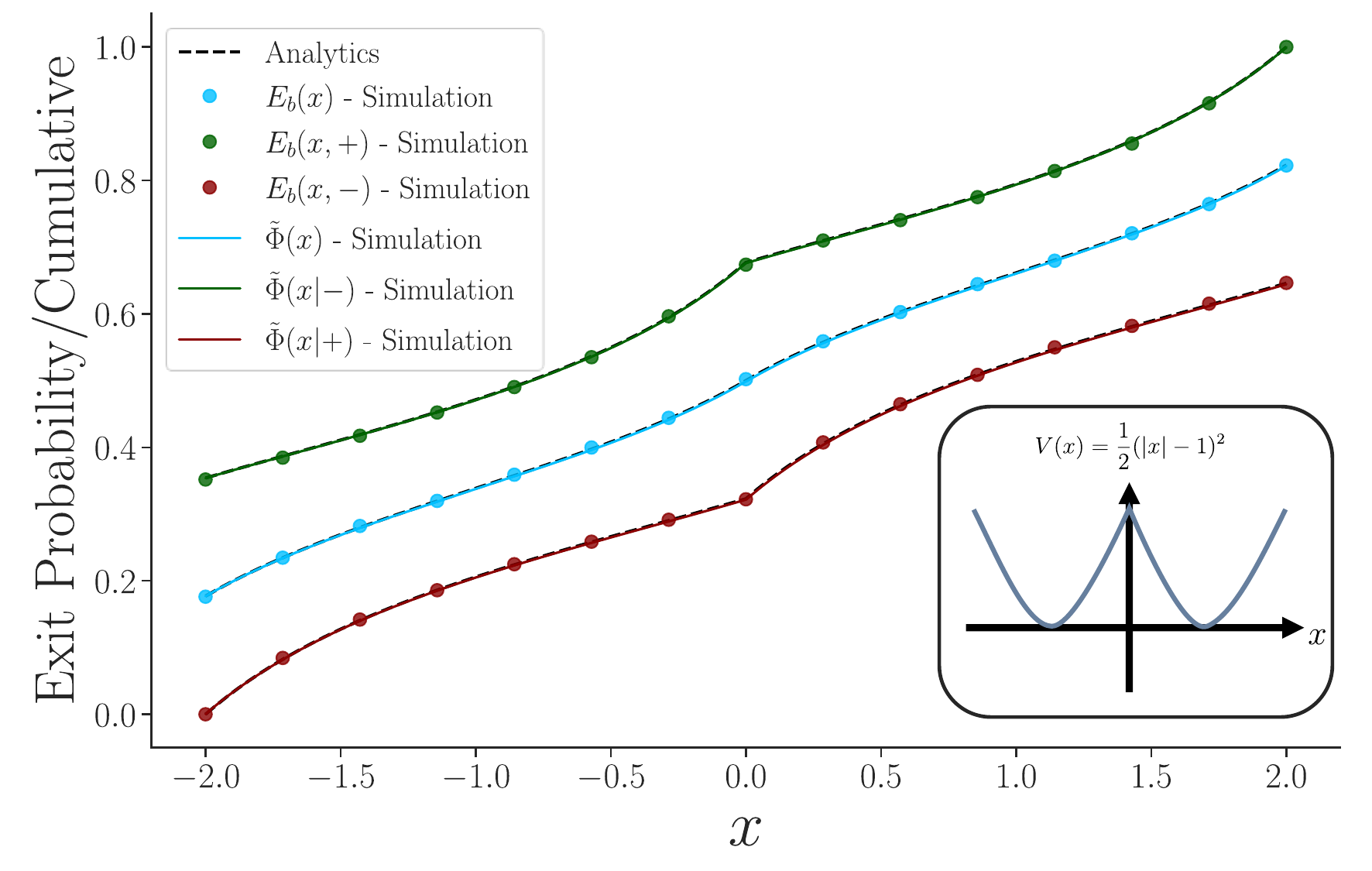}
    \end{minipage}
    \caption{This figure illustrates the duality relations $E_b(x) = \tilde{\Phi}(x)$ (\ref{duality1RTPstatio}) and $E_b(x,\pm) = \tilde{\Phi}(x|\mp)$ (\ref{duality2RTPstatio}) for a run-and-tumble particle (RTP). We compute the exit probability (resp. the cumulative distribution) of an RTP in various external potentials $V(x)$ (resp. $-V(x)$), with absorbing walls (resp. hard walls) located at positions $x=a$ and $x=b$. The solid lines show numerical simulations of the cumulative $\tilde{\Phi}$ while the dots are obtained from simulations of the exit probability $E_b$. The data from  $E_b$ and $\tilde{\Phi}$ overlap exactly. Explicit analytical solutions (dotted lines) are also plotted in dashed line. Throughout this analysis, we ensure that the condition $|F(x)|<v_0$ is satisfied.
    Top left: $V(x) = -\alpha\, x$, $\gamma = 1$, $\alpha = 0.5$, $v_0= 1$, $a=-1$, $b=1$.
    Top right: $V(x) = \alpha |x|$, $\gamma = 1$, $\alpha = 0.6$, $v_0= 1$, $a=-1$, $b=1$.
    Bottom left: $V(x) = \mu\, x^2/2$, $\gamma = 1$, $\mu= 1.9$, $v_0= 2$, $a=-1$, $b=1$.
    Bottom right: $V(x) = 1/2\, (|x|-1)^2$, $\gamma = 1$, $v_0= 2$, $a=-2$, $b=2$.}
\label{figureRTPdualitystatio}
\end{figure}

We also provide in Appendix \ref{appendix1} the explicit formulas when the RTP moves inside a linear potential and a double-well potential. In Figure \ref{figureRTPdualitystatio}, we compare our formulas to simulations and the agreement is perfect. For all these computations, we made sure that inside the interval $[a,b]$ the force verifies $|F(x)|<v_0$. 

\section{Distribution of the position of a run-and-tumble particle with hard walls}\label{cumulativeRTPstationarySection}
Let us now consider a seemingly completely different problem, namely the calculation of the stationary density of a run-and-tumble particle subjected to an external force $F(x)$, confined between two impenetrable walls at $x=a$ and $x=b$ (see Figure \ref{figureRTPwalls}). We again assume $|F(x)|<v_0$ for any $x$ in the interval $[a,b]$. The case of a general force is commented in Section \ref{generalforceRTP}. The solution without walls was found in \cite{DKM19} for an arbitrary $F(x)$, while the solution in the presence of walls but without the external force was presented in \cite{AngelaniHardWalls}. It is straightforward to combine both methods to obtain the solution to our problem.

Let us denote $P(x,+)$ and $P(x,-)$ the densities of an RTP in the $+$ and $-$ states, which are normalised such that $\int_a^b dx \left[P(x,+) + P(x,-)\right] = 1$.
The steady-state Fokker-Planck equations for these densities are given by
\begin{eqnarray}
&&\partial_x \left[\left(F(x)+v_0\right)P(x,+)\right] + \gamma\,  P(x,+) - \gamma\,  P(x,-) = 0\, , \\
&&\partial_x \left[\left(F(x)-v_0\right)P(x,-)\right] - \gamma\,  P(x,+) + \gamma\,  P(x,-) = 0\, .
\end{eqnarray}
Introducing $P(x)=P(x,+)+P(x,-)$ and $Q(x)=P(x,+)-P(x,-)$ we get
\begin{eqnarray}
&&\partial_x \left[F(x)\, P(x) + v_0\, Q(x)\right]  = 0 \label{eq1Density}\, ,\\
&&\partial_x \left[F(x)\, Q(x) + v_0\, P(x)\right] +2\gamma\,  Q(x) = 0 \, , \label{eq2Density}
\end{eqnarray}
as in \cite{DKM19}. The difference arises when considering the boundary conditions. Due to the persistent motion, the RTP may remain at either wall for a finite time. More precisely, the density $P(x,-)$ will have a finite mass $\kappa_a$ at $x=a$, and $P(x,+)$ will have a finite mass $\kappa_b$ at $x=b$.
Since $\kappa_a$ and $\kappa_b$ are stationary, the total current $J(x)=J(x,+)+J(x,-)$, where $J(x,\pm) = (F(x) \pm v_0)P(x,\pm)$, should vanish at the boundaries. In addition the probability current of a $+$ (resp. $-$) particle at $x=a$ (resp. $x=b$) arises entirely from a $-$ (resp. $+$) particle stuck at the wall which switches sign. Therefore one can write $J(a,+) = - J(a,-) = \gamma\,  \kappa_a$ and $J(b,+) = - J(b,-) = \gamma\,  \kappa_b $ which translates to 
\begin{eqnarray}
&&\left[v_0+F(a)\right]P(a,+) = \left[v_0-F(a)\right]P(a,-) = \gamma \, \kappa_a \, ,\label{leftedge}\\
&&\left[v_0+F(b)\right]P(b,+) = \left[v_0-F(b)\right]P(b,-) = \gamma \, \kappa_b \, ,\label{rightedge}
\end{eqnarray}
and implies
\begin{equation}
F(a)\, P(a)+v_0\,  Q(a) = 0 \quad , \quad 
F(b)\, P(b)+v_0\,  Q(b) = 0\, .
\end{equation}
Integrating \eqref{eq1Density} and taking this condition into account gives for all $x$
\begin{equation}
F(x)\, P(x) + v_0\,  Q(x) = 0\, ,
\end{equation}
which we can use to replace $Q(x)$ in \eqref{eq2Density}. We thus obtain the equation
\begin{equation}
\partial_x [(v_0^2-F(x)^2)\, P(x)] - 2\gamma\, F(x) P(x) = 0\, ,
\end{equation}
that is solved by
\begin{equation}
P(x) = \frac{1}{Z}\, \frac{2\gamma\, v_0}{v_0^2-F(x)^2} \exp \left[ 2\gamma \int_a^x dz\, \frac{F(z)}{v_0^2-F(z)^2} \right]\, ,
\label{exprPbulk}
\end{equation}
for any $x$ in $(a,b)$, with $Z$ a normalisation constant. This expression is the same as  the density of an RTP without walls derived in \cite{DKM19}. The difference comes from the normalisation and the presence of delta functions at the walls. The expressions for $P(x,\pm)$ can be easily deduced from this result
\begin{equation}
P(x, \pm) = \frac{1}{2}(P(x) \pm Q(x)) = \frac{1}{2} \left(1 \mp \frac{F(x)}{v_0} \right) P(x) = \frac{\gamma}{Z}\, \frac{1}{v_0 \pm F(x)}\, \exp \left[ 2\gamma \int_a^x dz\, \frac{F(z)}{v_0^2-F(z)^2} \right]\, .
\label{exprPpmbulk}
\end{equation}
Using (\ref{leftedge})-(\ref{rightedge}) we deduce
\begin{equation}
\kappa_a = \frac{1}{Z} \quad , \quad \kappa_b = \frac{1}{Z} \exp \left[ 2\gamma \int_a^b dz\, \frac{F(z)}{v_0^2-F(z)^2} \right] \, .\label{leftmass}
\end{equation}
The constant $Z$ is then fixed by the normalisation condition 
$\int_{a^+}^{b^-} dx\, P(x) + \kappa_a + \kappa_b = 1$. Thus the full expression for $P(x)$ is
\begin{eqnarray}
\fl 
    P(x) = \frac{1}{Z} \left( 2\gamma v_0 \frac{\exp \left[ 2\gamma \int_{a}^x du\, \frac{F(u)}{v_0^2 - F(u)^2} \right]}{v_0^2-F(x)^2} + \delta(x-a) + \exp \left[ 2\gamma \int_a^b du\, \frac{F(u)}{v_0^2-F(u)^2} \right] \delta(x-b) \right) \, , \nonumber \\
\fl     Z = 2\gamma v_0 \int_{a}^b dz\, \frac{\exp \left[ 2\gamma \int_{a}^z du\, \frac{F(u)}{v_0^2 - F(u)^2} \right]}{v_0^2-F(z)^2} + \exp \left[ 2\gamma \int_{a}^b du\, \frac{F(u)}{v_0^2 - F(u)^2} \right] + 1\, .
\end{eqnarray}
Let us now write the associated cumulative distribution. One has, for any $x$ in $[a,b]$,
\begin{equation}
\Phi(x) = \int_{a^-}^x dz\, P(z) = \frac{1}{Z} \left(2\gamma v_0 \int_{a}^x dz\, \frac{\exp \left[ 2\gamma \int_{a}^z du\, \frac{F(u)}{v_0^2 - F(u)^2} \right]}{v_0^2-F(z)^2} + 1 \right)\, ,
\label{cumulative}
\end{equation}
which is exactly the same result as \eqref{exitprobaexplicitRTP} for the probability to exit at $x=b$ when starting at $x$, but with an opposite force $-F(x)$. From Eq.~(\ref{exprPpmbulk}), we can express $P(x,\sigma)$ for $x\in [a,b]$ and $\sigma=\pm 1$ by adding Kronecker symbols $\delta_{\sigma, \pm}$ to take into account the finite probability masses of $-$ and $+$ at the walls. It gives
\begin{equation}
P(x,\sigma) = \frac{1}{Z} \left( \gamma \frac{\exp \left[ 2\gamma \int_{a}^x dy \frac{F(y)}{v_0^2 - F(y)^2} \right]}{v_0 + \sigma F(x)} + \delta_{\sigma,-}\delta(x-a) + \delta_{\sigma,+}\exp \left[ 2\gamma \int_a^b dy \frac{F(y)}{v_0^2-F(y)^2} \right] \delta(x-b) \right)\, .
\label{exprPpm_final}
\end{equation}
Integrating over $x$ and dividing by $P(\sigma)=\frac{1}{2}$ (the probability for the particle to be in the state $\sigma$ in the stationary state), we obtain the cumulative of the positions of the dual process conditioned on the internal state $\sigma$ for $x\in (a,b)$,
\begin{equation}
\begin{split}
&\Phi(x|\sigma) = \frac{1}{P(\sigma)} \int_{a^-}^x dz\, \, P(z, \sigma) = \frac{2}{Z} \left(\gamma \int_{a}^x dz\, \frac{\exp \left[ 2\gamma \int_{a}^z du\, \frac{F(u)}{v_0^2 - F(u)^2} \right]}{v_0 + \sigma F(z)} + \delta_{\sigma,-} \right)\label{cumulative_step} \\
&= \frac{2}{Z} \left(\gamma v_0 \int_{a}^x dz\, \frac{\exp \left[ 2\gamma \int_{a}^z du \frac{F(u)}{v_0^2 - F(u)^2} \right]}{v_0^2 - F(z)^2} - \sigma \gamma \int_{a}^x dz \frac{F(z)}{v_0^2 - F(z)^2} \exp \left[ 2\gamma \int_{a}^z du\, \frac{F(u)}{v_0^2 - F(u)^2} \right] + \delta_{\sigma,-}\right). \\
\end{split}
\end{equation}
One can then simplify by noticing that the integrand of the second integral is the derivative of $1/(2\gamma)\exp \left[2\gamma \int_{a}^z du\, \frac{F(u)}{v_0^2 - F(u)^2} \right]$. We thus finally obtain
\begin{equation}
\Phi(x|\pm) = \frac{1}{Z} \left(2\gamma v_0 \int_{a}^x dz\, \frac{\exp \left[ 2\gamma \int_{a}^z du\, \frac{F(u)}{v_0^2 - F(u)^2} \right]}{v_0^2 - F(z)^2} \mp \exp \left[ 2\gamma \int_{a}^x dz\, \frac{F(z)}{v_0^2 - F(z)^2} \right] + 1 \right)\, . 
\label{cumuPMRTP}
\end{equation}
The weights of the delta peaks in Eq.~(\ref{exprPpm_final}) are recovered from $\Phi(a^+|\sigma)=\kappa_a\, \delta_{\sigma,-}$ and $1-\Phi(b^-|\sigma)=\kappa_b\, \delta_{\sigma,+}$. In the next subsection we will relate $\Phi(x|\pm)$ to the exit probabilities $E_b(x,\pm)$. 

\section{Duality relation for a run-and-tumble particle in the stationary state}\label{SectionRTPDUAL}

The strong similarity between Eqs. \eqref{exprEpm_final} and \eqref{cumuPMRTP} suggests the existence of a relation between the exit probability and the cumulative distribution of an RTP in the presence of hard walls. We will now clarify this connection.

Let us consider again the process $x(t)$ defined in Eq.~(\ref{langeRTP}). We define its dual process $y(t)$ through the dynamics
\begin{equation} \label{defRTP_dual}
    \dot{y}(t)=-F(y) + v_0\, \tilde{\sigma}(t)\, ,
\end{equation}
where $\tilde{\sigma}(t)$ is a different realisation of the same telegraphic noise defined in Eq.~(\ref{telegraphicdynamics}). The process $y(t)$ describes the motion of a run-and-tumble particle subjected to the reversed force $-F(y)$. In addition, the process $y(t)$ has hard walls at $a$ and $b$. We denote by $\tilde{\Phi}(y)$ the cumulative distribution of the dual process $y(t)$. From Eq.~(\ref{cumulative}), we observe that one can link the exit probability of an RTP (\ref{exitprobaexplicitRTP}) to the cumulative of its dual via the following relation
\begin{equation}
    E_b(x)=\tilde{\Phi}(x)\, .
\label{duality1RTPstatio}
\end{equation}
One can also write a more general relation for the two states of the RTP relating Eq.~(\ref{exprEpm_final}) to Eq.~(\ref{cumuPMRTP}),
\begin{equation}
E_b(x,\pm) = \tilde \Phi(x|\mp)\, .
\label{duality2RTPstatio}
\end{equation}
Note that the derivation of the exit probability is simpler than the one of the cumulative, in particular because the boundary conditions for the density are more subtle. In Figure \ref{figureRTPdualitystatio}, we check numerically these relations for different forces $F(x)$ such that $|F(x)|<v_0$. 

Let us stress an important consequence of this relation, which is specific to active particles (in the absence of diffusion). On the one hand, an RTP with initial position at $x=a^+$ can still go away from the wall and eventually exit at $x=b$ as long as its velocity is positive at early times, and thus $E_b(a^+,+)$ is generally non-zero (and similarly $E_b(b^-,-)<1$). On the other hand, active particles tend to accumulate near hard walls, leading to delta peaks in the stationary density at $x=a$ and $x=b$ (where the RTP is in the $-$ and $+$ state respectively), such that $\Phi(a^+,-)>0$ and $\Phi(b^-,+)<1$. The identity \eqref{duality2RTPstatio} shows that this two phenomena are related, since $E_b(a^+,+)=\tilde \Phi(a^+,-)$ and $E_b(b^-,-)=\tilde \Phi(b^-,+)$ (see Fig.~\ref{figureRTPdualitystatio}).

The identity \eqref{duality2RTPstatio} is actually valid not only in the stationary state, but at any time $t$, provided that the initial conditions are chosen correctly. This is what we show in the next section.

\section{Duality relation for a run-and-tumble particle at finite time} \label{sec:finite_time}

In Section \ref{SectionRTPDUAL}, we have established the duality relation (\ref{duality2RTPstatio}) for RTPs in the stationary state, i.e, at infinite time.  It is natural to ask whether such a duality is also valid at finite time. Explicitly obtaining the exit probability, or the cumulative distribution of an RTP at finite time is highly non trivial, especially in the presence of an external force. However, without computing them, we can show that they obey the same differential equation with the same boundary and initial conditions.

Let us consider again the process $y(t)$ defined in \eqref{defRTP_dual} (still with hard walls at $a$ and $b$). In the following, we denote $\tilde F(x)=-F(x)$ the external force applied to $y(t)$, and we write with a tilde all the quantities related to $y(t)$. Previously we were only interested in the stationary distribution of this process, so that the initial condition did not matter. However, if we want to extend the duality to finite time, we need to specify the initial condition, for the position $y(0)$, but also for the state $\tilde \sigma(0)$. We will now show that the duality relation between $x(t)$ and $y(t)$ still holds at finite time if $y(0)=b$ and if $\tilde \sigma(0)=\sigma_0$ is drawn from its stationary distribution $\mathbb{P}(\sigma_0=\pm 1) = 1/2$.

We begin by writing the forward Fokker-Planck equation satisfied by the density of the position of $y(t)$, $\tilde P(y,t|\tilde \sigma(t) = \pm1;y(0)=b, \sigma_0)$, where the semicolon in the conditioning separates finite time conditions from initial conditions. We then deduce from it the differential equation verified by the cumulative distribution, defined as
\begin{equation}
\tilde \Phi(x,t|\tilde \sigma(t) = \pm1;y(0)=b) = \frac{1}{2} \int_{a^-}^x dy \sum_{\tilde \sigma_0=\pm 1}  \tilde P(y,t|\tilde \sigma(t) = \pm1;y(0)=b, \sigma_0)\, . 
\end{equation}
For convenience, in the rest of this section we will denote $\tilde P_\pm(y,t) = \tilde P(y,t|\tilde \sigma(t) = \pm1;y(0)=b, \sigma_0)$ and $\tilde \Phi_\pm(x,t|b) =\tilde \Phi(x,t|\tilde \sigma(t) = \pm1;y(0)=b)$, and we will often omit the arguments.

\begin{figure}[t]
\centering
    \begin{minipage}[c]{0.49\linewidth}
        \centering
        \includegraphics[width=1.\linewidth]{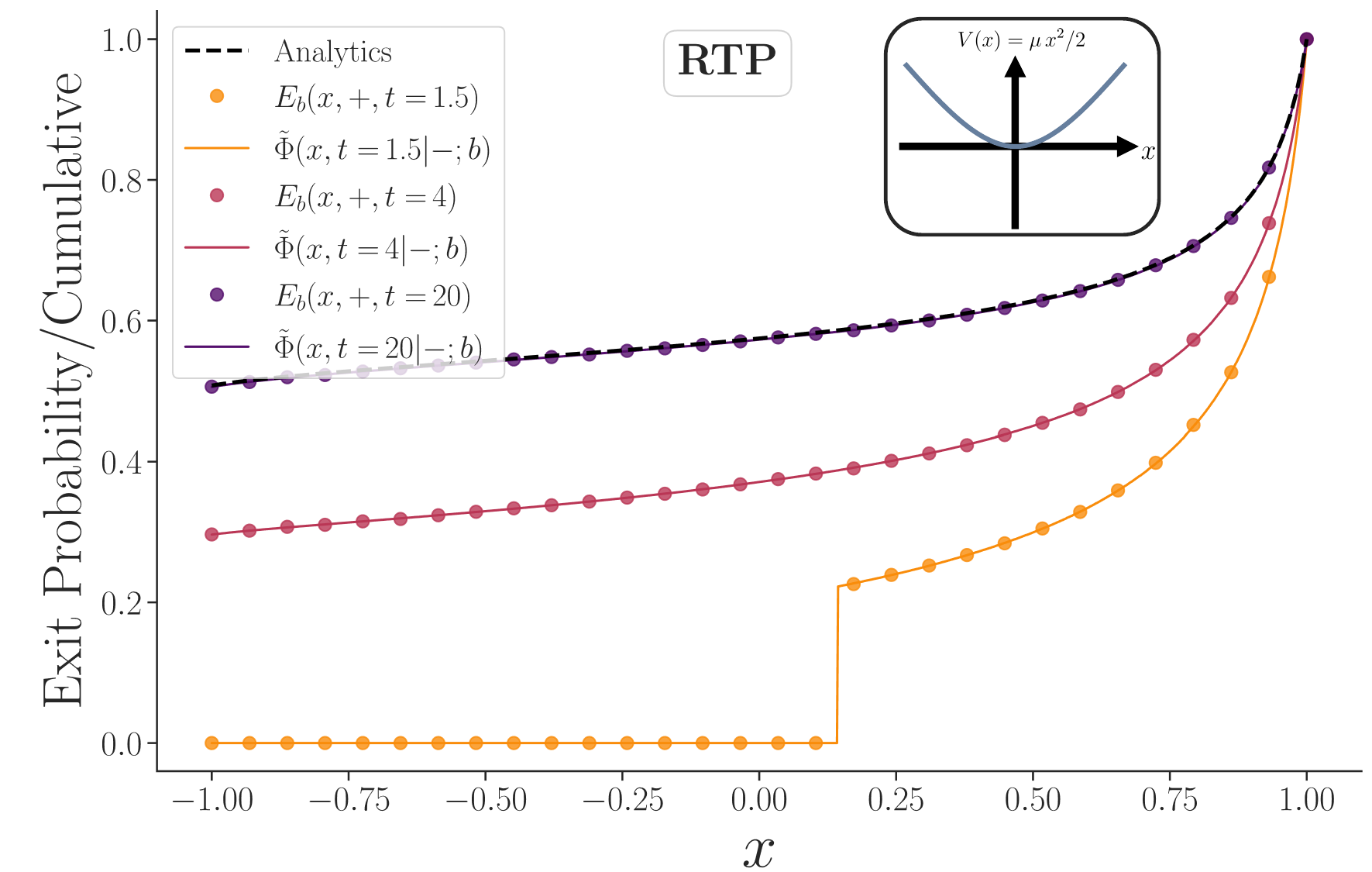}
    \end{minipage}
    \begin{minipage}[c]{0.49\linewidth}
        \centering
        \includegraphics[width=1.\linewidth]{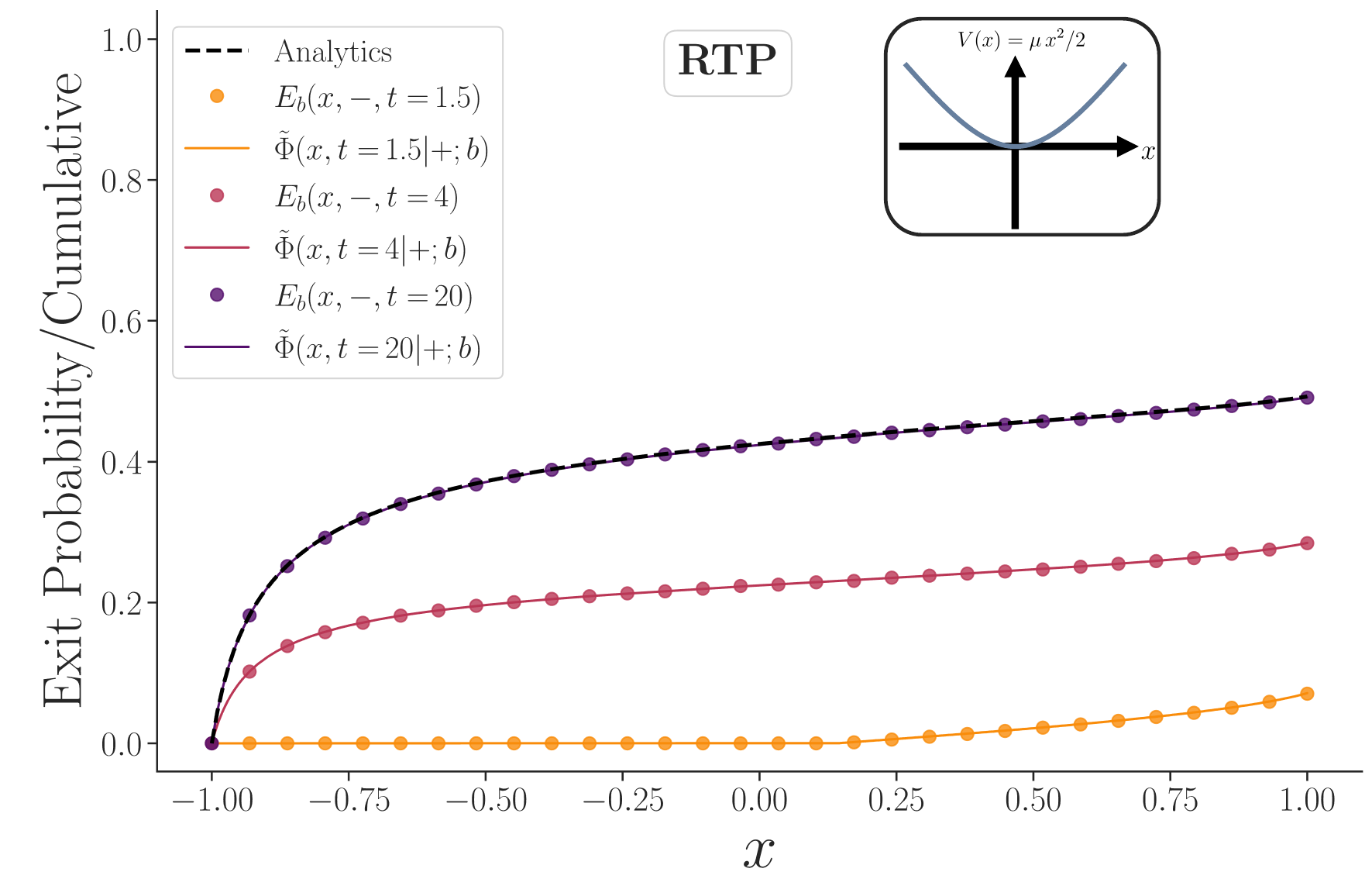}
    \end{minipage}
    \caption{Illustration of the duality relation \eqref{duality_finite_time} at finite time $t$ for an RTP in a harmonic potential. {\bf Right:} the dots represent the exit probability $E_b(x,+,t)$ in the presence of a potential $V(x)=\mu\,  x^2/2$, with absorbing walls at $a=-1$ and $b=1$, while the lines show the cumulative distribution of the dual $\tilde \Phi(x,t|-;b)$, with a potential $-V(x)$ and hard walls at $a$ and $b$. {\bf Left:} Same plot with the $+$ and $-$ particles exchanged. All the results were obtained by averaging over $10^6$ simulated trajectories, with parameters $\mu=1.9$, $v_0=2$ and $\gamma=1$. The dashed black lines show the analytic predictions for the stationary state.}
\label{figureRTPtime}
\end{figure}

We start by writing the Fokker-Planck equations  satisfied by the particle density 
\begin{eqnarray}
&&\partial_t \tilde P_+ = - \partial_x \left[(\tilde F(x)+v_0)\tilde P_+\right] - \gamma \tilde P_+ + \gamma \tilde P_-\, , \\
&&\partial_t \tilde P_- = - \partial_x \left[(\tilde F(x)-v_0)\tilde P_-\right] + \gamma \tilde P_+ - \gamma \tilde P_-\, .
\end{eqnarray}
Since we choose the initial distribution of $\sigma_0$ to be the stationary distribution $p_{st}(\tilde \sigma=\pm 1)=\frac{1}{2}$, the distribution of $\tilde \sigma(t)$ is independent of time. Using $\tilde P_\pm(x,t)= \partial_x \tilde \Phi_\pm(x,t)$ we write
\begin{eqnarray}
&&\partial_x [-\partial_t \tilde \Phi_+ -(\tilde F(x)+v_0)\partial_x \tilde \Phi_+ - \gamma\tilde \Phi_+ + \gamma \tilde \Phi_-] = 0\, , \label{deriveephi1}\\
&&\partial_x [-\partial_t \tilde \Phi_- -(\tilde F(x)-v_0)\partial_x \tilde \Phi_- + \gamma \tilde \Phi_+ - \gamma \tilde \Phi_-] = 0\, .\label{deriveephi2}
\end{eqnarray}
We now want to integrate equations (\ref{deriveephi1}) and (\ref{deriveephi2}) over $x$. As we have already seen in Sec.~\ref{cumulativeRTPstationarySection}, the boundary conditions need to be treated with care.
Indeed the density $\tilde P_-$ (resp. $\tilde P_+$) has a finite mass at $x=a$ (resp. $x=b$), given by $\tilde \Phi_-(a^+,t|b)=\kappa_a(t)$ (resp $1-\tilde \Phi_+(b^-,t|b)=\kappa_b(t)$). However, since there is no mass of $+$ particles at $x=a$ nor of $-$ particles at $x=b$, we have $\tilde \Phi_+(a^+,t|b)=1-\tilde \Phi_-(b^-,t|b)=0$. The current of $+$ particles at $x=a^+$ is thus generated entirely by the $-$ particles at $x=a$ which switch sign. Thus the boundary condition at $x=a^+$ reads $(\tilde F(a^+)+v_0)\tilde P_+(a^+,t|b) = \gamma \kappa_a(t)$, or equivalently,
\begin{equation}
(\tilde F(a^+)+v_0)\partial_x\tilde \Phi_+(a^+,t|b) = \gamma \tilde \Phi_-(a^+,t|b)\, ,
\end{equation}
and similarly at $x=b^-$, exchanging $+$ and $-$ particles. Thus we have
\begin{eqnarray}
\!\!\!\!\!\!\!\!\!\!\!\!\!\!\!\!\!\! 0&=&-\partial_t \tilde \Phi_+(a^+,t|b) -(\tilde F(a^+)+v_0) \partial_x \tilde \Phi_+(a^+,t|b) - \gamma \tilde \Phi_+(a^+,t|b) + \gamma \tilde \Phi_-(a^+,t|b)\, ,\\
\!\!\!\!\!\!\!\!\!\!\!\!\!\!\!\!\!\! 0&=&-\partial_t \tilde \Phi_-(b^-,t|b)-(\tilde F(b^-)-v_0) \partial_x \tilde \Phi_-(b^-,t|b) + \gamma \tilde \Phi_+(b^-,t|b) - \gamma \tilde \Phi_-(b^-,t|b)\, .
\end{eqnarray}
Therefore there is no integration constant when integrating \eqref{deriveephi1}-\eqref{deriveephi2} and we can write, for any $x$ in $(a,b)$,
\begin{eqnarray}
&&\partial_t \tilde \Phi_+ = -(\tilde F(x)+v_0) \partial_x \tilde \Phi_+ - \gamma \tilde \Phi_+ + \gamma \tilde \Phi_-\, , \\
&&\partial_t \tilde \Phi_- = -(\tilde F(x)-v_0) \partial_x \tilde \Phi_- + \gamma \tilde \Phi_+ - \gamma \tilde \Phi_-\, .
\end{eqnarray}
These are exactly the backward Fokker-Planck equations \eqref{timeBFPplus}-\eqref{timeBFPminus} for the exit probabilities of the process $x(t)$ , $E_b(x,-,t)$ and $E_b(x,+,t)$, with an external force $F(x) = -\tilde F(x)$. In addition $\tilde \Phi_\pm(x,t|b)$ and $E_b(x,\pm,t)$ satisfy the same boundary conditions
\begin{eqnarray}
&&E_b(a^+,-,t)=\tilde \Phi_+(a^+,t|b)=0\, , \\
&&E_b(b^-,+,t)=\tilde \Phi_-(b^-,t|b)=1\, ,
\end{eqnarray}
as well as the same initial conditions
\begin{equation}
E_{b}(x,\pm,0)=\tilde \Phi_\mp(x,0|b)=\begin{cases} 0 \text{ if } x<b \;, \\ 1 \text{ if } x=b \;. \end{cases}
\end{equation}
Therefore, denoting again $\tilde \Phi_\mp(x,t|b)$ as $\tilde \Phi(x,t|\mp ;b)$, we can write the generalized version of~\eqref{duality2RTPstatio}\footnote{We admit here without proof the unicity of the solution of the PDEs with these initial and boundary conditions.},
\begin{equation} \label{duality_finite_time}
E_{b}(x,\pm,t)=\tilde \Phi(x,t|\mp ;b)
\;.
\end{equation}
This relation is illustrated through numerical results in Fig. \ref{figureRTPtime}, for an RTP in a harmonic potential. Of course the finite time version of \eqref{duality1RTPstatio} is also valid. Let us also note that the relation \eqref{duality_finite_time} (both at finite and infinite time) has an equivalent for the exit probability at $x=a$, $E_a(x,v,t)$, which can be immediately deduced by symmetry (one simply needs to replace $\tilde \Phi$ by $1-\tilde \Phi$ and to choose $y(0)=a$).

\section{Comments on the duality relation of an RTP subjected to a general force in the stationary state }\label{generalforceRTP}

For this section we focus again on the stationary state. We present an extension of the results of Section~\ref{SectionRTPDUAL} to encompass a broader range of forces that do not satisfy the condition $|F(x)|<v_0$. This extension allows for forces with multiple turning points, i.e. points such that $F(x) = \pm v_0$, within the interval $[a, b]$, as well as regions where the force either falls below $-v_0$ or exceeds $v_0$ for certain values of $x$ in the same interval (see \cite{Velocity_RTP} where a similar situation is considered for the stationary distribution of an RTP). Unlike in Sec.~\ref{cumulativeRTPstationarySection}, the stationary state of $y(t)$ now depends on the initial condition. However, we will show that, using the same initial condition as for the finite time result, namely $y(0)=b$, the duality relation \eqref{duality2RTPstatio} still holds.

\begin{figure}
  \centering
    \includegraphics[width=0.49\textwidth]{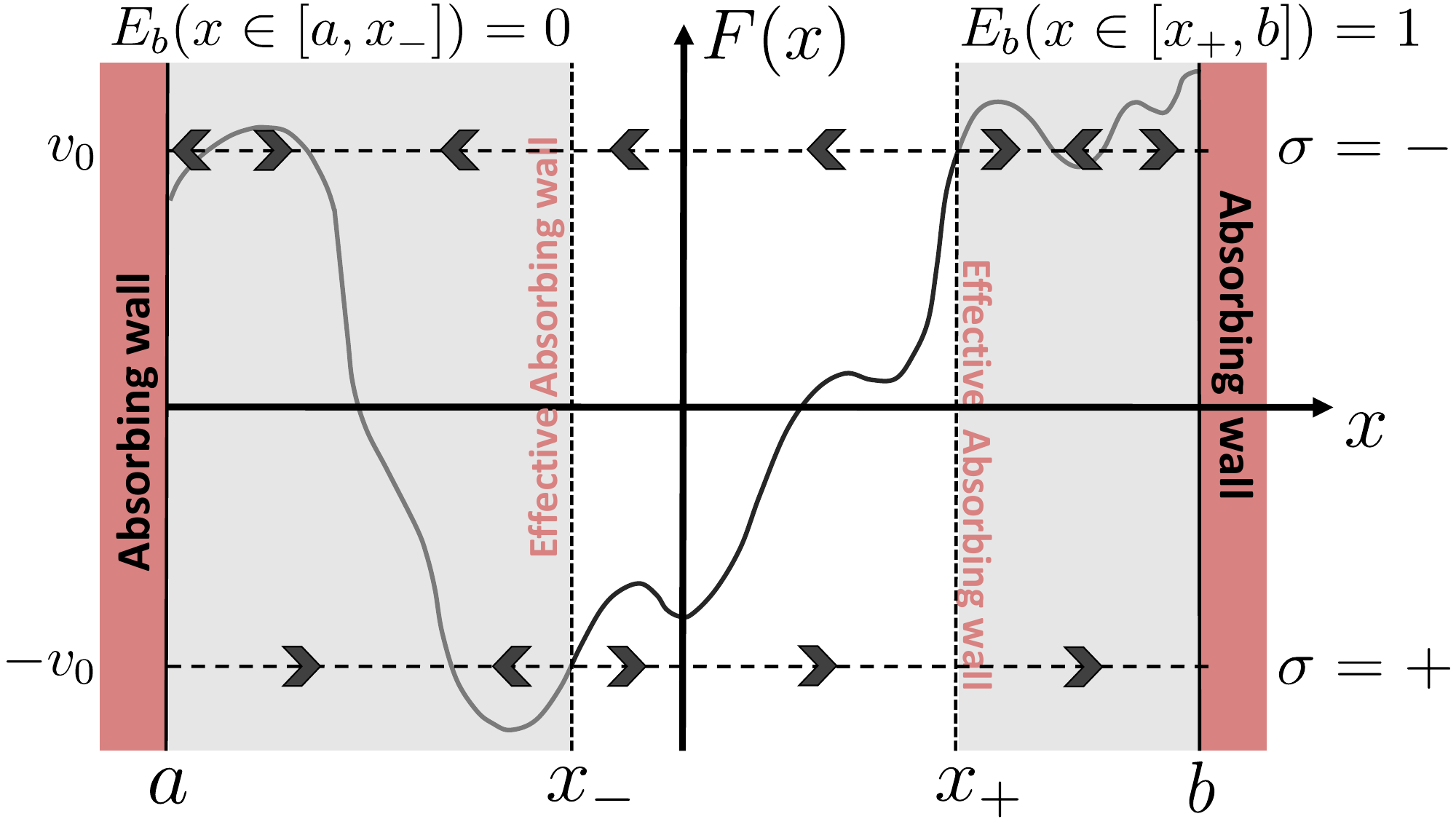}
    \includegraphics[width=0.49\textwidth]{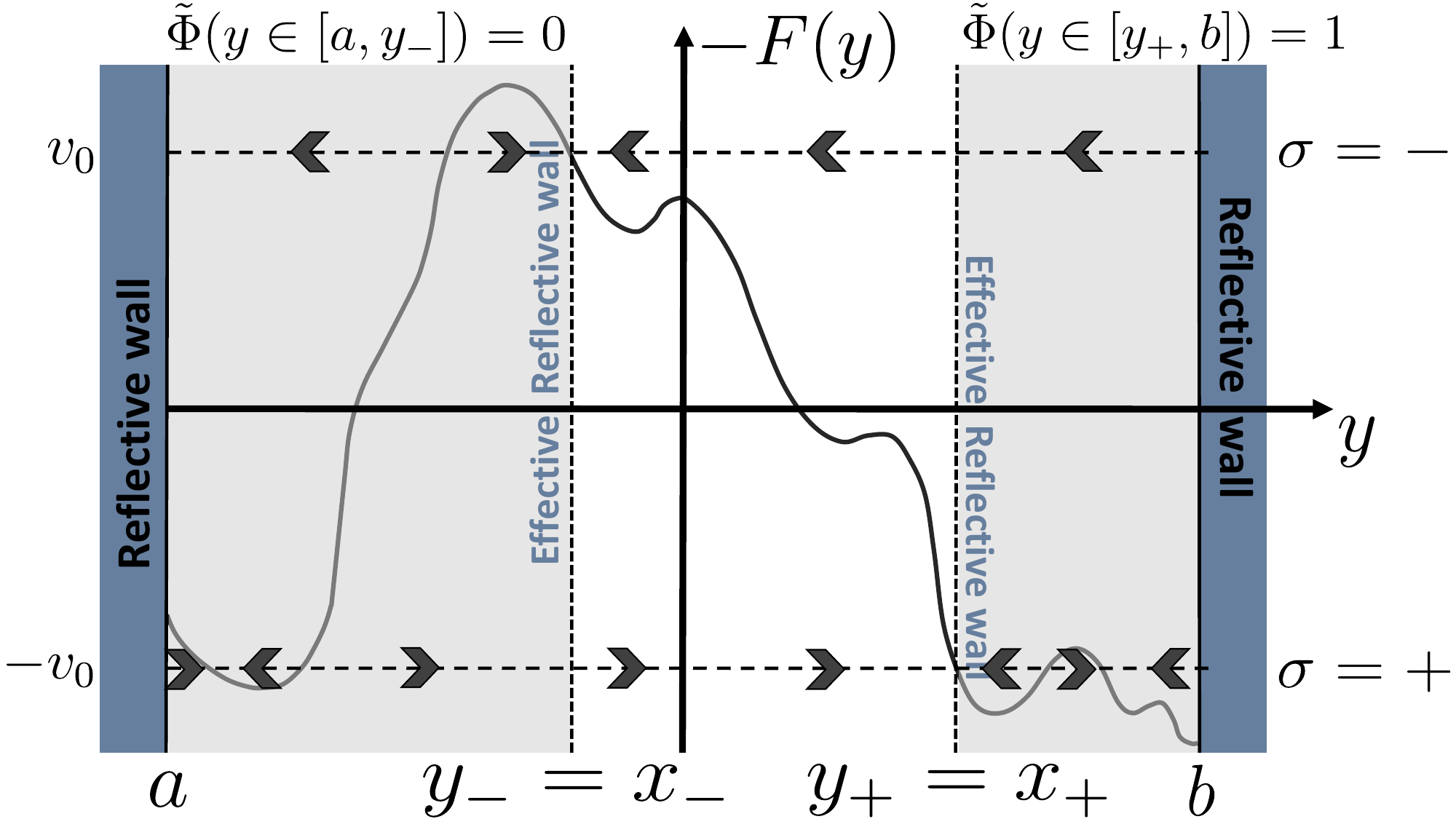}
  \caption{We show here an illustration of the statements of Section~\ref{generalforceRTP} which aim to generalize results from Section~\ref{SectionRTPDUAL} for a force $F(x)$ that can take any value in $\mathbb{R}$. We are interested in the calculation of the exit probability of an RTP (right panel), and in the cumulative distribution of its dual (left panel). We represent the velocity of the negative state of the RTP on the top dotted line, while the velocity of the positive state is shown on the bottom dotted lines of the plots. If the velocity of the particle is positive, the arrow points to positive values of the abscissa, while if it is negative, it points to the opposite.}
  \label{fig:two_subfigs_force}
\end{figure}

We begin by examining the exit probability for an RTP described by the process $x(t)$ defined in \eqref{langeRTP}. There are two absorbing walls located at $x=a$ and $x=b$. To do so, we introduce the following definition:
\begin{equation}
    x_- = \ {\rm sup} \left(\{x \in [a,b] \ | \ F(x)  \leq -v_0\} \cup \{a\} \right)\; .
\end{equation}
Both states of the RTP have negative (or null) velocity at $x_-$ implying that $E_b(x\in[a,x_-]|\pm)=0$. Consequently, if the particle initiates its motion inside $]x_-, b]$ and reaches the position $x_-$, it will remain within the interval $[a, x_-]$ at all subsequent times. Naturally, it will be unable to exit at $x = b$. As a result, in terms of the exit probabilities (at infinite time) $E_b(x, \pm)$, the system behaves as if there were an absorbing wall positioned at $x_-$. Let us now define
\begin{equation}
    x_+ = \ {\rm inf} \left(\{x \in [x_-,b] \ | \ F(x) \geq v_0\} \cup \{b\} \right)\; .
\end{equation}
At the position $x_+$, $\pm$ states of the RTP have a positive (or null) speed, and we also have $F(x)>-v_0$ for all $x\in [x_+,b]$. Hence, if the particle reaches $x_+$ from its left, it is guaranteed to eventually exit at $x = b$ so that $E_b( x\in[x^+,b])=1$. Therefore, if we are only interested in the exit probability at infinite time, we can consider that the absorbing wall at $x=b$ is actually replaced by an effective one at $x_+$. For an illustration, see the left panel of Figure~(\ref{fig:two_subfigs_force}).

Now, let us consider the dual process $y(t)$ defined in \eqref{defRTP_dual}.
At $t=0$, the dual process begins at position $y=b$, and there are two hard walls at $a$ and $b$. As mentioned earlier, unlike the case where $|F(y)|<v_0$ holds everywhere, the stationary state of this dual process will depend on its initial condition. In this context, we introduce the following definition
\begin{equation}
    y_- = \ {\rm sup} \left(\{y \in [a,b] \ | \ -F(y) \geq v_0\} \cup \{a\} \right)\; .
\end{equation}
At position $y_-$, the negative state of the dual RTP has zero speed, while the positive state has a positive velocity. Consequently, as the particle starts its motion at $y=b$, it will always remain at positions such that $y \geq y_-$.  In particular, this leads to $\tilde \Phi(y\in [a,y_-])=0$. Therefore, introducing a hard wall at $y_-$ will not change the stationary distribution of positions. Similarly, we define
\begin{equation}
    y_+ = \ {\rm inf} \left(\{y \in [y_-,b] \ | \ -F(y)  \leq -v_0\} \cup \{b\} \right)\; .
\end{equation}
At position $y_+$, the negative state of the particle has a negative speed, while the positive state has zero speed. Furthermore, for all $y \in [y_+,b]$, we have $-F(y)<v_0$ and thus, in a finite amount of time, the particle that started at $b$ will reach $y_+$. Consequently, we will have $\tilde \Phi(y \in [y_+,b])=1$. Therefore, if our objective is solely to determine the stationary distribution of the dual process, we can assume the existence of a hard wall at $y_+$.

At this point, it becomes clear that the definitions of $x_-$ and $x_+$ coincides with the one of $y_-$ and $y_+$. In Figure~\ref{fig:two_subfigs_force}, we provide an illustration of these statements. One can then apply the results of Section \ref{exitprobaRTPsubsection} and Section \ref{cumulativeRTPstationarySection} on the interval $(x_-,x_+)$ replacing $a$ with $x_-+\epsilon$ and $b$ with $x_+-\epsilon$ and taking $\epsilon \to 0$ (some integrals diverge in this limit, but the divergences compensate leading to a well-defined finite limit). When $x_-$ or $x_+$ is strictly inside $(a,b)$, there is no discontinuity of $E_b$ and $\tilde \Phi$ at this point. For the exit probability this is because a particle at $x_- + \epsilon$ will have zero total velocity if it is in the $+$ state (since $F(x_-)=-v_0$), and thus it will not be able to escape, leading to $E_b(x_- + \epsilon,+)=0$ (and similarly $E_b(x_+ - \epsilon,-)=1$). For the stationary distribution, this is because it requires an infinite time for the particle to reach the walls (since for instance $-F(x_-)=v_0$), and thus there is no delta peak at the walls (hence $\tilde\Phi(x_-+\epsilon|-)=0$ and $\tilde\Phi(x_+-\epsilon|+)=1$). In both cases it is as if the particle inside the interval $(x_-,x_+)$ does not ``feel'' the presence of the walls. For the stationary density, one thus recovers the results of \cite{DKM19}.

It is important to note that, in the scenario where turning points are present, we do not necessarily have $E_a(x,\pm)=1-E_b(x,\pm)$. However, a similar reasoning can be made to obtain the exit probability at $a$. In this case, we define:
\begin{equation}
\begin{split}
    &x_+ = \ {\rm inf} \left(\{x \in [a,b] \ | \ F(x) \geq v_0\} \cup \{b\} \right)\; , \\
    &x_- = \ {\rm sup} \left(\{x \in [a,x_+] \ | \ F(x) \leq -v_0\} \cup \{a\} \right)\; , 
    \end{split}
\end{equation}
or similarly for $y_-$ and $y_+$.

As a conclusion, all the duality results of Sec.~\ref{SectionRTPDUAL} remain true if there are points such that $|F(x)|>v_0$, although one has to be more careful when computing both $E_b$ and $\tilde \Phi$. In particular one should use the proper initial condition for $y(t)$, i.e. $y(0)=b$.

\section{Discussion} \label{discussion}

We have introduced a duality relation for run-and-tumble particles connecting the exit probability with the cumulative distribution of the position in the presence of hard walls, at any time. In the stationary state, we obtained explicit expressions for both quantities. In a future work \cite{LongPaper}, we will show how to derive this relation in a much more general setting, that encompasses not only the most studied models of active particles, but also other stochastic processes which are of interest in physics. 

The duality identity~(\ref{duality_finite_time}) can be thought of as a sort of time reversal. For every trajectory of a particle starting at $x$ in the presence of a potential $V(x)$ and exiting at $b$ before or at time $t$, one can construct a trajectory of a dual particle starting at $b$ with a reversed potential $-V(x)$ and located inside $[a,x]$ at time $t$. This becomes clearer when deriving the duality relation Eq.~(\ref{duality_finite_time}) in discrete-time as done in~\cite{LongPaper}.

This duality relation has several practical applications. From an analytical point of view, deriving one of the two quantities gives directly access to the other as a byproduct. Sometimes, one of them is easier to derive than the other, as we have seen with the example of the RTP. If instead one seeks to compute these quantities from numerical simulations or experiments, there might also be situations where one quantity is simpler to obtain than the other. For instance if one is interested in the exit probability at infinite time, one a priori needs to observe many trajectories starting from every position $x$ in $[a,b]$. Instead, one can compute the stationary distribution between hard walls by observing a single trajectory and averaging over time (assuming that the system is ergodic). This could be of particular relevance in experiments, where it is often easier to observe a single long trajectory (see e.g. Ref.~\cite{singletraj}).

\bigskip

\noindent\textbf{Acknowledgments}

\noindent We thank P. Le Doussal, S. N. Majumdar and G. Schehr for useful comments and discussions.
We also thank J. Klinger and B. De Bruyne for interesting discussions in the early stage of the project.

\appendix
\renewcommand\thesection{\Alph{section}}

\section*{Appendix}

\section{Explicit formulas for the exit probabilities of an RTP inside a linear potential and a double-well potential}\label{appendix1}

\textit{Linear potential:} The linear potential $V(x)= \alpha\, |x|$ leads to a force $F(x) = -\alpha\, \text{sign}(x)$. This force is constant up to the sign of $x$. To have $|F(x)|<v_0$ in $[a,b]$, we impose $|\alpha|<v_0$. We also fix the location of the walls:  $a=-1$, and $b=1$. For positive values of $x$, i.e. $0\leq x \leq1$, one has 
\begin{eqnarray}
    E_b(x,\pm) = \frac{1}{Z} \left[ 1 \pm e^{\frac{2\gamma\alpha(x - 1)}{v_0^2-\alpha^2}} + \frac{v_0}{\alpha} \left(1 - 2e^{-\frac{2\gamma\alpha}{v_0^2-\alpha^2}} + e^{\frac{2\gamma\alpha(x - 1)}{v_0^2-\alpha^2}}\right) \right] \, ,
\end{eqnarray}
while for negative values of $x$, i.e. $-1\leq x\leq 0$, one has 
\begin{eqnarray}
    E_b(x,\pm) = \frac{1}{Z} \left[ 1 \pm e^{-\frac{2\gamma\alpha(1 + x)}{v_0^2 - \alpha^2}} + \frac{v_0}{\alpha} \left(1 - e^{-\frac{2\gamma\alpha(1 + x)}{v_0^2 - \alpha^2}}\right) \right] \, .
\end{eqnarray}
The normalisation constant reads \begin{eqnarray}
    Z = 2 \left[1 + \frac{v_0}{\alpha} \left(1 - e^{-\frac{2\gamma\alpha}{v_0^2-\alpha^2}}\right) \right]\, .
\end{eqnarray}
\textit{Double-well potential:} In this scenario, we examine a potential described by $V(x) = \frac{\mu}{2} \left(|x| - x_0\right)^2$, which gives rise to a force  $F(x) = \mu (x_0 - |x|)\, \text{sign}(x)$. We constrain the location of the walls such that $b<1+v_0$ and $a>-1-v_0$ in order to have $|F(x)|<v_0$ in $[a,b]$. For positive values of $x$, the exit probabilities are given by (here we fix $x_0=1$ and $\mu=1$ to simplify the expressions)
\begin{flalign}
&E_b(x,\pm)= \frac{1}{Z} \left\{ 1 \pm \left(\frac{(1 + a)^2 - v_0^2}{v_0^2 - (x - 1)^2}\right)^\gamma + \frac{2\gamma}{v_0}\, \left(1- \frac{(1 + a)^2}{v_0^2}\right)^\gamma
\left[2 \, _2F_1\left(\frac{1}{2}, 1 + \gamma, \frac{3}{2}, \frac{1}{v_0^2}\right)  \right. \right. \nonumber \\
&\left.\left. - (1 + a) \, _2F_1\left(\frac{1}{2}, 1 + \gamma, \frac{3}{2}, \frac{(1 + a)^2}{v_0^2}\right) + (x - 1) \, _2F_1\left(\frac{1}{2}, 1 + \gamma, \frac{3}{2}, \frac{(x - 1)^2}{v_0^2}\right)\right] \right\}\, . &&
\end{flalign}
For negative values of $x$, they read
\begin{flalign}
&E_b(x,\pm) = \frac{1}{Z} \left\{ 1\pm \left(\frac{(1 + a)^2 - v_0^2}{(1 + x)^2 - v_0^2}\right)^{\gamma}+ \frac{2\gamma}{v_0}\, \left(1- \frac{(1 + a)^2}{v_0^2}\right)^\gamma \right. \nonumber\\
    & \left. \times \left[(1 + x) \, _2F_1\left(\frac{1}{2}, 1 + \gamma, \frac{3}{2}, \frac{(1 + x)^2}{v_0^2}\right) - (1 + a) \, _2F_1\left(\frac{1}{2}, 1 + \gamma, \frac{3}{2}, \frac{(1 + a)^2}{v_0^2}\right)\right] \right\} \, . &&
\end{flalign}
 The normalisation is
 \begin{flalign}
Z =1 + \left(\frac{(1 + a)^2 - v_0^2}{(1 - b)^2 - v_0^2}\right)^\gamma 
+ \frac{2\gamma}{v_0}\, \left(1- \frac{(1 + a)^2}{v_0^2}\right)^\gamma
    \left[2\, _2F_1\left(\frac{1}{2}, 1 + \gamma, \frac{3}{2}, \frac{1}{v_0^2}\right) \right. \quad \quad \quad \nonumber \\
    \left.- (1 + a) \, _2F_1\left(\frac{1}{2}, 1 + \gamma, \frac{3}{2}, \frac{(1 + a)^2}{v_0^2}\right) + (b - 1) \, _2F_1\left(\frac{1}{2}, 1 + \gamma, \frac{3}{2}, \frac{(b - 1)^2}{v_0^2}\right)\right] \, . &&
\end{flalign}

\end{document}